# RAN Enablers for 5G Radio Resource Management

D. M. Gutierrez-Estevez, Ö. Bulakci, M. Ericson, A. Prasad, E. Pateromichelakis, J. Belschner, P. Arnold, G. Calochira

*Abstract*— **This paper presents the description of several key RAN enablers for the radio resource management (RRM) framework of the fifth generation (5G) radio access network (RAN), referred to as building blocks of the 5G RRM. In particular, the following key RAN enablers are discussed: i) interference management techniques for dense and dynamic deployments, focusing on cell-edge performance enhancement; ii) dynamic traffic steering mechanisms that aim to attain the optimum mapping of 5G services to any available resources when and where needed by considering the peculiarities of different air interface variants (AIVs); iii) resource management strategies that deal with network slices; and iv) tight interworking between novel 5G AIVs and evolved legacy AIVs such as Long-term Evolution (LTE). Evaluation results for each of these key RAN enablers are also presented.**

## I. Introduction

The explosive growth in capacity and coverage demands emerged the evolution of traditional Radio Access Networks (RANs) towards highly densified and heterogeneous deployments as foreseen in some fifth generation (5G) scenarios. The 5G radio access technology, through the support for extreme mobile broadband, massive machine-type communication and ultra-reliable communication, is expected to address the significant increase in data rate demands that network operators are expecting during the coming years. Due to the wide range of frequency bands used and the need to tailor the air interface parameters depending on the frequency band, 5G landscape is expected to consist of multiple air interface variants (AIVs), which could include evolved legacy technologies, e.g., long-term evolution-advanced (LTE-A) air interface as one component. Since 5G needs to support a wide range of diverse use cases and requirements such as extreme mobile broadband with 1000 times higher capacity, ultra-reliability of 99.999 % and low-latency of less than 1 ms over the air interface, it is expected that the network would be optimized for the target use case and the associated requirements, as well.

In this paper, we present the highlights of several key RAN enablers developed in the context of the METIS-II project, namely interference management, dynamic traffic steering, resource management for network slices, and tight interworking between 5G and LTE AIVs [1]. Firstly, interference management in current cellular networks has been extensively studied in literature. 5G networks pose novel challenges to the design of interference mitigation techniques such as tailoring its operation to the dynamic topologies envisioned in such networks, i.e., performing activation/deactivation of nomadic access nodes (NNs) to attain on-demand network densification for coverage and capacity enhancement while coping with the momentarily changing interference conditions both on the uplink and downlink [2]. In addition, 5G RAN is expected to operate on various bands (below and above 6 GHz) and support various 5G services with wide range of requirements. Secondly, a more dynamic mechanism for traffic steering is required as a key RAN enabler in 5G networks, which could be complemented by a dynamic definition and enforcement of quality of service (QoS). In legacy networks, traffic steering was considered as a key enabler for load balancing and improving user throughput [3]. Various mobility based traffic steering strategies for LTE-A heterogeneous networks were studied in [4], where each user is connected to the best layer that can serve them. Thirdly, resource management for network slicing enables the support of the network slicing concept [5] with respect to RRM when multiple slices are served on shared resources. For this purpose, the new logical entity Air interface agnostic Slice Enabler (AaSE) is defined. It introduces a control loop for the Service Level Agreement (SLA) associated with a network slice. Based on QoS class adaptation the AaSE influences user specific data flows to meet SLA requirements of multiple slices. Finally, the state-of-the-art integration between previous different AIV systems such as 3G and 4G is based on the traditional hard inter-RAT handover 0. The major drawbacks with inter-RAT hard handover e.g. between 3G and 4G are the rather long delay and service interruption as well as the low reliability. A tighter integration with evolved LTE may therefore be crucial in order to ensure ultra-high reliability and extreme bit rates in a 5G system.

A description of each of the aforementioned building blocks along with their evaluation results will be presented in the subsequent sections: Section II presents interference management, Section III dynamic traffic steering, Section IV resource management for network slices, and Section V tight

D. M. Gutierrez-Estevez is with Samsung Electronics R&D Institute UK, South Street, Staines-upon-Thames, TW18 4QE, United Kingdom (email: d.estevez@samsung.com).

Ö. Bulakci and E. Pateromichelakis are with Huawei Technologies GRC, Riesstr. 25, 80992, Munich, Germany (e-mail: {oemer.bulakci, emmanouil.pateromichelakis}@huawei.com).

M. Ericson is with Ericsson Research, Laboratoriegränd 11, 977 53 Luleå, Sweden (email: marten.ericson@ericsson.com).

A. Prasad is with Nokia Bell Labs, P.O. Box 226, 00045 Nokia Group, Finland (email: athul.prasad@nokia-bell-labs.com).

J. Belschner and P. Arnold are with Deutsche Telekom AG, Deutsche-Telekom-Allee 7, 64295 Darmstadt, Germany (email: {paul.arnold, jakob.belschner}@telekom.de).

G. Calochira is with Telecom Italia, Via Guglielmo Reiss Romoli, 274, 10148 Turin, Italy (email: giorgio.calochira@telecomitalia.it).

integration with evolved legacy AIVs. Finally, conclusions are drawn in Section VI.

## II. INTERFERENCE MANAGEMENT

The interference management building block is a key element of the agile RRM framework. Enhanced and extended functionalities must be added in 5G systems to deal with highly dynamic networks and topologies requiring a high level of flexibility. To present in this paper an exhaustive list of interference management techniques for 5G is out of the scope of this paper. Fig. 1 shows a sample scenario where different types of techniques relevant to METIS-II are employed simultaneously according to the spatio-temporal needs of the network; these include techniques using dynamic topologies, advanced modulation schemes, or interference orthogonalization. In this section, we summarize two proposed techniques addressing the dynamism of the 5G topology and the required flexibility based on advanced modulation schemes but by no means we imply that that this would be an exhaustive description of the 5G interference building block, which would indeed require a greater number of interference avoidance and mitigation techniques.

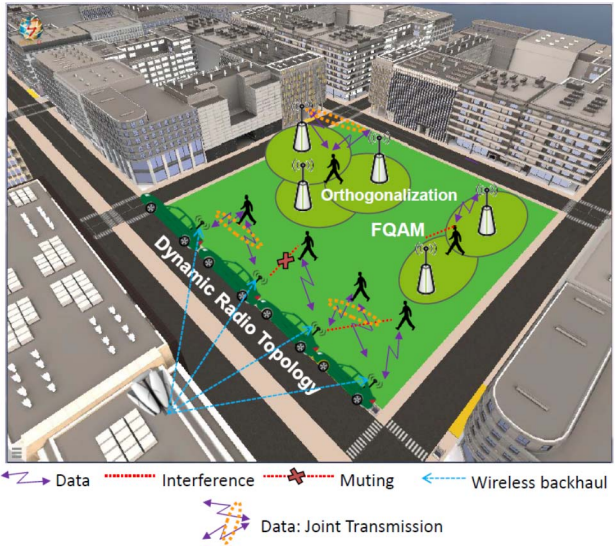

Figure 1. Interference management technologies for 5G.

### A. Interference management in dynamic topologies

A key aspect of the interference management building block is to provide UE-centric interference management in heterogeneous UDNs by means of selecting overlays of access nodes that can serve users individually, given their diverse service requirements. On top of that, coordinated resource allocation and joint transmission will be applied adaptively based on the backhaul conditions, the load constraints and the service type. Here, we provide a case study for a hotspot area and a 5G RAN consisting of a number of NNs under a macro-cell umbrella. In particular, we consider a dynamic network topology comprising such non-static access nodes, which emerges as a promising notion enabling flexible network deployment and new services.

The key interference management mechanisms which are applied are Joint Transmission (JT) between the access links of NNs (i.e., between NNs and users) when it is possible. The selection of candidate users for JT is based on the difference of their channel measurements (reference signal received power, RSRP) from serving and neighboring NNs. Given the number of users with low channel quality, a number of resource blocks (RBs) are reserved for JT, and resource allocation between different NNs is performed. For the rest of users, coordinated scheduling is applied, where dynamic frequency partitioning (or muting of resources for some NNs) is performed.

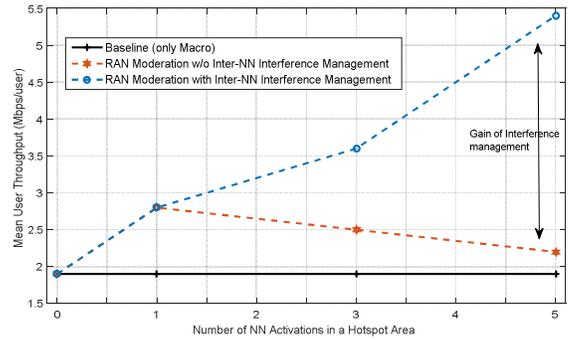

Figure 2. Mean user throughput for different NN activations.

The results are demonstrated in Fig. 2. The curves show the mean user throughput for the cases when we activate NNs and also when we perform interference management on top of that. It is shown that via the activation of NNs, up to 50% mean user throughput can be achieved in case of one active NN. However, as can be seen in Fig. 5 for more than one activation in a hotspot area, the achievable gain decreases when more NNs are activated, which is due to interference from neighboring NNs. Therefore, interference management is crucial particularly when the network density increases. As further shown, adaptive interference coordination and cooperation mechanisms can substantially improve mean user throughput in dynamic radio topologies.

### B. Interference management based on advanced modulation for cell edge users

It has been shown that inter-cell interference (ICI) in conventional cellular networks employing orthogonal frequency division multiple-access (OFDMA) with Quadrature Amplitude Modulation (QAM) tends to approach a Gaussian distribution [7]. Furthermore, it has also been proven that the worst-case additive noise in wireless networks with respect to the channel capacity has a Gaussian distribution [8]. However, recent studies show that combining QAM with frequency-shift keying (FSK) into what is termed as frequency and quadrature-amplitude modulation (FQAM) can be advantageous to change the pattern of ICI into non-Gaussian when applied at interfering cells, hence improving the performance of low SINR users in neighboring cells [9]. To increase the flexibility of FQAM-based interference management techniques, FQAM can be orthogonally 'partitioned' along different dimensions of the radio resources, namely space, frequency, and time, as follows: i) for a spatial split of resources, only certain interfering beams are selected for employment of FQAM; ii) for a frequency-based split of resources, a flexible FQAM resource pool is negotiated among base stations; and iii) already established time-based procedures (e.g., ABS) can be enhanced with FQAM-based subframes to effectively improve the data rate of the edge users experiencing heavy interference.

Herein, we focus on the spatial dimension, i.e., on combining FQAM with beamforming techniques such that base stations inducing beamformed interference cause as little performance degradation as possible to neighboring cells. For that, an algorithm is employed that detects highly interfered users and coordinates the base stations via similar techniques to 3GPP's coordinated scheduling/coordinated beamforming to extract the gains of FQAM [10]. Fig. 3 left shows that the 95% available transmission rates, i.e., the lower 5% of the rate distribution curve, can still be significantly improved by applying FQAM only to those users experiencing high level of interferences, and QAM to the rest. This is because the available rate mainly depends on the users in low SINR regime, and FQAM can precisely improve throughput dramatically in low SINR regime. The figure also shows the average transmission rates still lower than FQAM. However, applying FQAM does not affect average transmission rates that significantly because although low-SINR users are experiencing improved performance, high-SINR users do experience significant improvements with FQAM.

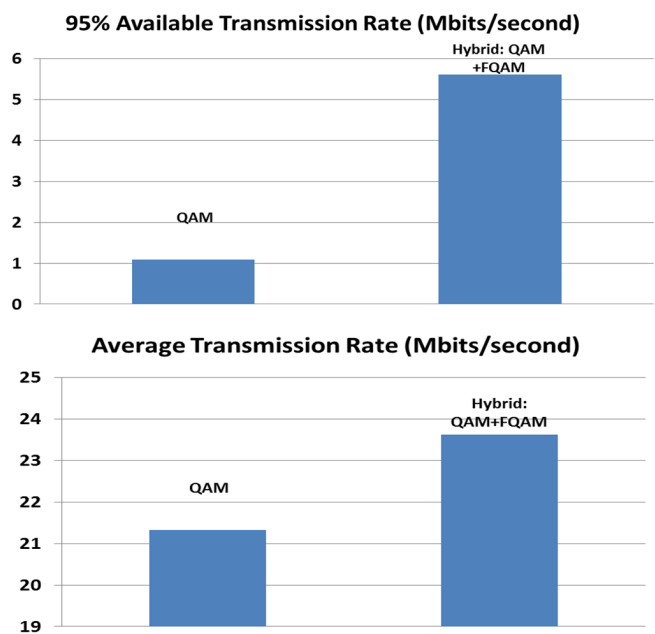

Figure 3. Transmission rates for interference management based on FQAM frequency partitioning.

## III. DYNAMIC TRAFFIC STEERING

5G networks have unique requirements such as ultra-reliability, low-latency and high-capacity, which requires the system to execute functions and operate on a very fast timescale, as compared to legacy radio access technologies. For ultra-reliable machine type communication, data rates are not the key factor for optimization, but packet latency and link reliability are essential. But for use cases such as virtual reality, high data rates are also essential, apart from the reliability aspects. This makes functions such as dynamic traffic steering, which traditionally is considered to be an asynchronous function, with relatively less constraints on the operational speed. Dynamic traffic steering on a synchronous, time transmission interval (TTI) timescale can achieve several optimizations in the network. In this work, we discuss how such mechanisms can achieve energy savings and reduced delay in the network. A detailed evaluation of these mechanisms in terms of latency reduction and energy efficiency is presented in [12]. We consider the possible centralization of higher layer RAN functions in a logical entity called Access Network-Outer (AN-O) layer, and the lower-layer RAN functions in the AN-Inner (AN-I) layer. A similar approach is currently being studied in 3GPP as well, for the 5G / new radio related studies [11], where the AN-O and AN-I layers are called central and distributed units respectively.

For the multi-AI dynamic traffic steering concept presented in [12], the key consideration is to enable the RAN to steer the traffic over the multiple active AIs, in a synchronous time scales, depending on the real-time feedback from the AN-I layers. This is essential due to the relative link unreliability of the 5G networks, deployed in higher frequency bands. In order to enable this, the RAN is required to have more control over enforcing the QoS policies that it receives from the core network. Here it is assumed that these functions would be present at the AN-O layer. This would enable the AN-O layer to interact with the user gateways present in the core network for changing the end point of the traffic, as well as the AN-O layer to efficiently deliver the traffic to the end users. This will enable the network to remove resource reservations at the 5G-BSs as soon as a link failure is detected, thereby re-farming the resources for other ultra-reliable users, while steering the traffic towards other active links. This enables the efficient use of radio resources at the AN-I layer, while also ensuring that the QoS targets of the end user is met. The main limitation in this aspect with current systems is that for guaranteed bit rate traffic the RAN has limited control over removing resource reservation required for the dedicated bearer, which would be essential in 5G, in order to support the new requirements and use cases.

Dynamic traffic steering is essential to achieving network energy savings in 5G networks. Considering the avoidance of always-on signals in 5G, the RAN has the ability to enter and leave energy saving state for the duration of a few TTIs, if the load conditions are suitable for entering such an energy efficient state. The traffic steering for energy efficiency concept, considers the AN-O layer steering the traffic over multiple AN-I layer BS, depending on the real-time traffic load of the system. We consider the application of advanced transmission schemes such as JT, along with the capabilities available in the AN-O layer, to enable the optimal activation and use of 5G-BS, thereby achieving energy savings. Based on the evaluations done in [12], it was shown that the proposed dynamic traffic steering mechanisms can provide significant energy saving gains in the network.

Here we have discussed the use of dynamic traffic steering in order to enable a diverse set of 5G requirements such as reliability, latency, high capacity and energy efficiency. The need for having a functional decomposition of RAN functions into centralized and distributed units for achieving these requirements is also discussed. While traffic steering is one of the logical functions that can be located in the AN-O layer, it is an important one due to the ability to achieve key 5G requirements.

As an example, in a heterogeneous environment where systems operating at millimeter wave (mmW) and traditional bands co-exist, a proper mechanism to manage resources and cope with interference in mmW-bands can be introduced. The idea is to focus on a pre-emptive geometrical-based interference analysis (PGIA) that is able to determine, prior to the establishment of a new transmission link, a set of mutually interfering mmW transmission links (where incumbent and new links are grouped) allowing the network to implement a suitable resource partitioning mechanism (at AN-I level) of the identified set or take other alternative measures (e.g. traffic steering by establishing a transmission link on a lower frequency) at AN-O level. In particular, the proposed solution allows to limit transmission collisions (intended here as transmissions creating such mutual interference with neighbor transmission links so as to make the communication impossible) and to limit the subsequent signaling overhead aimed to solve the problem. Preliminary results depicted in Fig. 4 show that PGIA coupled with a simple resource sharing mechanism can significantly reduce the number of interfered links as the number of concurrent mmW links in 1 km$^2$ increase (in this exemplary analysis, a link is considered interfered with a C/I below 12 dB [13]). Without PGIA, the average percentage of interfered mmW links rises over 95% as the number of concurrent links grows to 200, while with a PGIA and resource sharing mechanism, the average percentage of interfered mmW links is capped around a very low 2.5%.

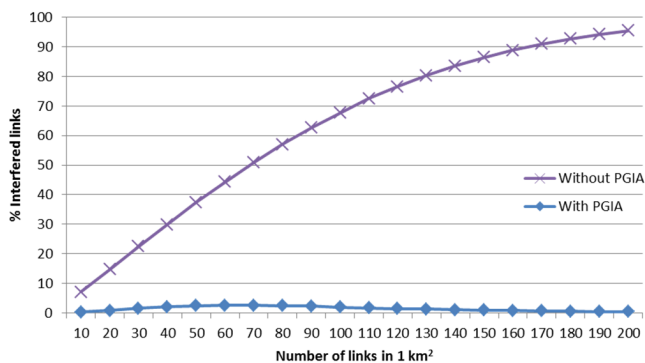

Figure 4. Average % of interfered links as functions of the number of concurrent links in 1 square km, without and with PGIA.

## IV. RESOURCE MANAGEMENT FOR NETWORK SLICES

The concept of network slicing is a key enabler for the huge variety of 5G services. It is based on the idea of running multiple logical networks as virtually independent business operations on a common physical infrastructure [5]. This extends the relatively static principle of network sharing [14] as outlined in the following.

With respect to RM, especially the management of the scarce radio resources is a critical issue. Thus, pooling and sharing these resources among the logical networks (the so called network slices) in an efficient manner is the main target.

The operator of the physical network infrastructure guaranties a certain network quality for each network slice. This is defined in the so called SLA. For example, a data rate of 1 Mbit/s and a maximum delay of 20 ms could be guaranteed for any data flow within one network slice whereas a second network slice has different guaranties. An SLA is often combined with a temporal component (e.g. that the guaranties have to be fulfilled in 99% percent of time) and a penalty that applies in case of SLA violations. In addition to the SLAs, each data flow can have dedicated QoS requirements.

Resource management (RM) for network slices is responsible of allocating the resources in a way that the SLAs of all network slices are fulfilled. It therefore fulfills the requirements agreed in 3GPP standardization [15] with respect to network slicing, such as RAN awareness of slices and RM between slices.

There are different approaches to implement RM for network slices with different levels of complexity. The basis for allocating resources in a slice aware manner is monitoring the current status of the network slices with respect to their SLAs. This could take place at a new entity of the RAN, e.g. an access controller as the Access Network-Outer (AN-O) layer used for dynamic traffic steering. The entity has to be aware of the existing network slices and their SLAs, as well as which data stream belongs to which network slice. This can be realized through signaling from the core network.

The enforcement of the network slice specific requirements happens with the help of existing QoS mechanisms of the 5G air interface variants. Based on the outcome of the SLA monitoring, the QoS Class Identifiers (QCIs) of the individual data streams are adjusted. If, for example, the SLA of a network slice guaranties a data rate of 1 Mbit/s per data stream, any data stream could be mapped to a corresponding QCI class. This mapping is a dynamic process which is supposed to solve conflicts between network slices in a way that all SLAs can be fulfilled.

Fig. 5 visualizes this process. The basis for a slice aware RM is that data flows from the core network are tagged with either information on the corresponding network slice or the corresponding SLA. An entity called Air interface agnostic Slice Enabler (AaSE) is responsible for monitoring and enforcing SLAs. As stated before, this could be part of an access controller. Based on the information from the core network and from the AIVs (e.g. from AIV specific schedulers), AaSE monitors the SLA status. An enforcement of SLAs happens by adapting the QoS classes of individual data streams. For example, a data stream from a network slice with high data rate guarantees can be configured to have a QoS class of a specific AIV with a guaranteed bitrate. For monitoring the SLA status, AaSE reads QoS KPIs of the AVIs. A feedback to the core network (whether SLAs are currently fulfilled) is important to monitor SLA status also there as well as to trigger network changes in case of constant SLA violations.

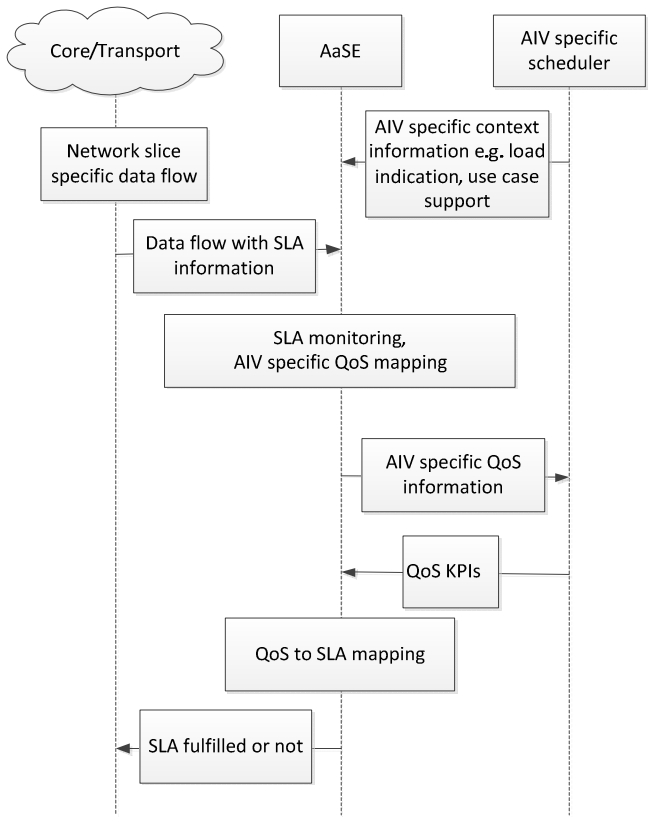

Figure 5. Implementation option for RM for network slicing.

Related ongoing work is also considered within other 5GPPP projects under the Horizon 2020 framework [16].

## V. Tight Integration with Evolved Legacy AIVs

5G is expected to operate in a wide range of frequency bands, probably using also very high frequency bands compared to 4G. This implies, for example, lower diffraction and higher outdoor-to-indoor penetration losses, which means that signals will have more difficulties to propagate around corners and penetrate walls. Also, the initial deployment of 5G will be rather spotty. Together with requirements from uMTC of ultra-reliably connection and extreme user bitrates of xMBB this motivates a more tight integration with legacy AIVs such as evolved LTE. This section evaluates the performance of such a tighter integration using a common Packet Data Convergence Protocol (PDCP) layer for both evolved LTE and 5G AIV.

The first evaluated concept is a fast user plane data (UP) switch at the (common) Packet Data Convergence Protocol (PDCP) layer. It is assumed that the control plane (CP) is using "dual connectivity" with LTE and 5G, while the UP is switched at PDCP level to either LTE or 5G. If the CP is connected to both the LTE node and the 5G node, no signaling is required and the UP switch may be almost instantaneous. Also, we assume a common S1 core network / radio access network (CN/RAN) interface for LTE and 5G, referred to as S1* herein. This means that no extra CN/RAN signaling is needed for a UP switch. The fast UP switch can be based on normal handover measurements such as RSRP. This is also used for the simulations shown below. Note that since the CP is active in both LTE and 5G, the reliability of the connection should increase compared to normal hard handover. Adding and deleting connections to new nodes may be based on LTE dual connectivity mechanism, i.e., based on the best connection for the user equipment (UE), namely downlink or uplink, but it can be also based on the load of the nodes or other triggers.

A second concept to investigate is when both UP and CP are connected to both LTE and 5G (similar to "dual connectivity" in LTE) and the UP data is aggregated (or split) at PDCP layer. Also for this solution we assume an S1* CN/RAN interface for LTE and 5G. This means that no extra CN/RAN signaling is needed to add or delete a secondary node. An alternative to the dual connectivity solution is to use the medium access control (MAC) layer for aggregation, as in CA for LTE. In this case, the scheduler can then use resources in an optimal way, at least if the UE is configured and able to send measurement information about all carriers (i.e. both LTE and 5G carrier). However, measurements and signaling to support this should also be possible to develop for the dual connectivity solution (still using PDCP as aggregation/split layer).

A benefit to use the PDCP layer to aggregate or split the data is the likely similarity between the PDCP layer for LTE and 5G, while the MAC layers may be rather different. Thus, using the PDCP layer will probably require less standardization efforts.

A drawback of having multiple flows of the CP is the increased overhead. Another potential drawback of a dual connectivity solution may be that a multi-connectivity solution for 5G might use a lower layer for aggregation (such as MAC layer). So, coordination between multi-connectivity within 5G together with dual connectivity on higher layer with LTE might require rather different signaling and solutions.

The above tight integration concepts have been evaluated using a system-level simulator. The evaluated concepts are the hard handover (HH), fast UP switch (FS) of the UP and the dual connectivity (DC) concepts. The LTE and 5G AI nodes are co-sited and the frequency bands investigated are 2 GHz for LTE and 15 GHz for 5G. The difference between LTE and the 5G AI is shorter transmission time interval (TTI) for 5G. The 5G AI has a TTI of 0.2 ms instead of 1 ms for LTE, as well as fewer sub-bands due to longer sub-carrier spacing. In the simulation environment there are 7 base stations (BSs) with 3 sectors each and inter-site distance (ISD) of 500 m. The radio channel model is the 3GPP Case 1 Urban Macro (UMa) channel model 0 where the attenuation constant is modified according to the carrier frequency. Fig. 6 shows the worst user, i.e. the 10%-ile user throughput vs. load for dual connectivity, hard handover and fast UP switch. The stand-alone 5G AI is used for comparison. It uses 15 GHz frequency and a bandwidth of 40 MHz (in contrast to 20+20 MHz for the tight integration cases).

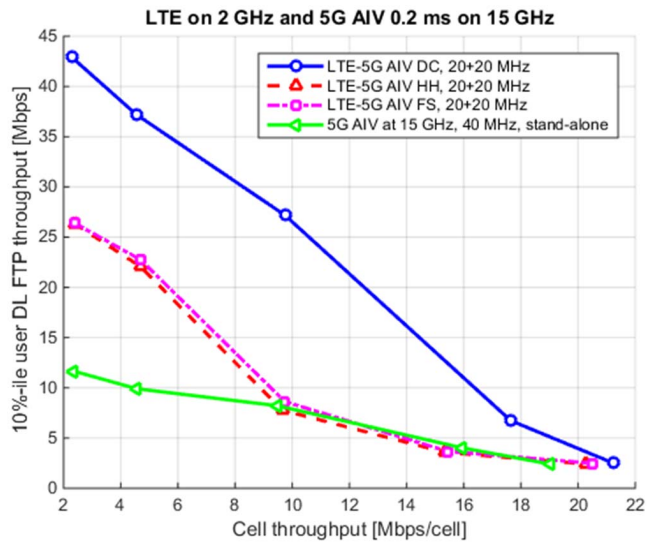

Figure 6. 10%-ile user throughput vs. load for dual connectivity (DC), hard handover (HH) and fast user plane switch (FS).

The dual connectivity concept shows the best performance, around 300% higher user throughput at low load compared to the stand-alone 5G AI case and around 100% higher compared to fast UP switch and hard handover cases. The difference in performance between hard handover and fast UP switch is small, even though hard handover has an interruption delay of 300 ms when a hard handover is performed compared to no delay at all for the fast switch. The reason for the small difference is due to the fact that there are very few hard handovers in this scenario and therefore the performance for hard handover is not affected very much.

VI. CONCLUSION

This paper has described some key RAN enablers of the 5G RRM, including interference management, dynamic traffic steering, resource management for network slices, and tight interworking between 5G and LTE. As backed by the corresponding results and analyses, these enablers constitute key building blocks that target the novel 5G aspects of diverse service requirements, overall air interface comprising multiple AIVs, dynamic radio topology and novel communication modes.

ACKNOWLEDGMENT

This work has been performed in the framework of the H2020 project METIS-II co-funded by the EU. The views expressed are those of the authors and do not necessarily represent the project. The consortium is not liable for any use that may be made of any of the information contained therein.